\documentclass[12pt,a4paper]{article}
\usepackage{standalone}
\usepackage[british]{babel}
\usepackage[hyphens]{url}
\usepackage{datetime}
\usepackage{amsmath}

\usepackage{ifthen}
\usepackage[utf8]{inputenc}
\usepackage[T1]{fontenc}
\usepackage{lmodern} 
\usepackage{eso-pic}
\usepackage{enumerate}
\usepackage{paralist}
\usepackage{longtable}
\usepackage{multirow}
\usepackage{float}
\usepackage{titleps}
\usepackage{rotating}
\usepackage{filecontents}
\usepackage{bibentry}
\usepackage{graphicx}
\usepackage{caption}
\usepackage{subcaption}
\usepackage{pdfpages}
\usepackage{pdflscape}
\usepackage{pgf}
\usepackage{tikz}
\usepackage{soul}
\makeatletter
\newcommand*{\textlabel}[2]{%
  \edef\@currentlabel{#1}
  \phantomsection
  #1\label{#2}
}
\makeatother

\usepackage[top=2.54cm,bottom=1.9cm,left=3.17cm,right=3.17cm]{geometry}
\usepackage[justification=justified,singlelinecheck=false]{caption}
\usepackage[sort&compress]{natbib}
\bibpunct{(}{)}{;}{a}{,}{,}

\renewcommand{\refname}{\normalsize References}
\newsavebox{\boxtext}
\def\hide#1{\savebox{\boxtext}{#1}}
\newdateformat{intldate}{\THEDAY\ \monthname[\THEMONTH] \THEYEAR}

\def\cf{{\it cf.}}

\begin{document}
\sloppy

\bibpunct{[}{]}{,}{n}{,}{,}
\def\sec#1{\section{#1}\vspace{-12pt}}
\def\subsec#1{\subsection{\normalfont{\bf #1}\vspace{-6pt}}}
\renewcommand{\refname}{\vspace{-60pt}\section{\textsc References}\vspace{-12pt}}
\footskip=15pt
\parindent 28pt
\parskip 12pt

\def\qual{qualitative}
\def\Qual{Qualitative}
\def\quant{quantitative}
\def\Quant{Quantitative}
\def\tm{TM}

\def\quest#1{\textcolor{green}{#1}}
\def\rev#1{\highlight[yellow!50!white, fill opacity=0.3, draw=red, very thick]{#1}}
\def\excerpt#1#2{%
 \begin{quotation}
 \noindent {\it #1} (#2)
 \end{quotation}
}

\def\subject{An experiment exploring the theoretical and methodological challenges in developing a semi-automated approach to analysis of small-N \qual\ data}

\author{
  Sandro Tsang, PhD\\
  Peoples Open Access Education Initiative\\
34 Stafford Road\\Manchester \ M30 9ED\\United Kingdom\\
  \texttt{skf.tsang[at]gmail.com}
}
\title{\LARGE{\subject}\\[12pt]\normalsize{\textsc{A Preprint}}}
\clearpage\maketitle
\thispagestyle{empty}

\begin{abstract}
This paper experiments with designing a semi-automated qualitative data analysis (QDA) algorithm to analyse 20 transcripts by using freeware.  Text-mining (TM) and QDA were guided by frequency and association measures, because these statistics remain robust when the sample size is small. The refined TM algorithm split the text into various sizes based on a manually revised dictionary. This lemmatisation approach may reflect the context of the text better than uniformly tokenising the text into one single size. TM results were used for initial coding.  Code repacking was guided by association measures and external data to implement a general inductive QDA approach.  The information retrieved by TM and QDA was depicted in subgraphs for comparisons. The analyses were completed in 6--7 days. Both algorithms retrieved contextually consistent and relevant information. However, the QDA algorithm retrieved more specific information than TM alone. The QDA algorithm does not strictly comply with the convention of TM or of QDA, but becomes a more efficient, systematic and transparent text analysis approach than a conventional QDA approach.  Scaling up QDA to reliably discover knowledge from text was exactly the research purpose.  This paper also sheds light on understanding the relations between information technologies, theory and methodologies.
\end{abstract}

\begingroup
\raggedright
\leftskip=28pt\rightskip=28pt
\noindent {\bf Keywords:} information retrieval, computation and language, multi-methods, qualitative data analysis, text-mining, philosophy of science, graph theory
\par
\endgroup


\pagebreak

\sec{Introduction}

Various research domains have been experimenting with algorithms which let human input be dynamically included in automated processes of knowledge discovery from textual data \citep{Winnenburgetal2008}.  The motivation is that humans are generally more accurate than text-mining (TM) in discovering knowledge, but TM can assist in scaling up the process \citep{Winnenburgetal2008}.  The research focus of this topic is often on analysis of big data.  Investigating this topic with textual data collected from few sources/participants (small-N) is also crucial, but draws little research attention.  It has implications for quickening dissemination of \qual\ evidence.  Medical practitioners need current best qualitative evidence to deliver medicine tailored to serve each patient as a unique individual \citep{Tonelli2018}.  \Qual\ research (QR) usually involves collecting and analysing unstructured textual material \citep{Kelle1997}.  The aim is to achieve depth of understanding, so the data are usually small-N data \citep{Palinkas2015}.  \Qual\ data analysis (QDA) can enable gaining a deep understanding of text \citep{Kelle1997,Wiedemann2013}.  It is a labour-intensive process \citep{Atkissonetal2016}.  It is not an efficient enabler for obtaining results to keep up with the pace of the growth of medical knowledge, which is estimated to double every 73 days by 2020 \citep{Densen2011}.  Integrative applications of TM and QDA will be a viable and relatively efficient solution.  Especially, TM facilities have been integrated with certain computer-assisted \qual\ data analysis software (CAQDAS) \citep{Atkissonetal2016,Eversetal2011,Kelle1997,Wiedemann2013}.  CAQDAS has been widely adapted for QDA, but largely for data management \citep{Kelle1997,Wiedemann2013}.  For \qual\ researchers, research philosophies, designs and QDA cannot be seen as separate entities \citep{GreenhalghTaylor1997,GreenBritten1998, Kelle1997, NobleSmith2015,Thomas2006,Wiedemann2013}.  Automating QDA requires rethinking of the relations between information technology (IT), theory and method (or methodology).  A consensus about this topic is yet to be reached \citep{Eversetal2011}.

TM and QDA operate on seemingly conflicting philosophies, but TM is seen as a complement of QDA and epistemologically compatible with \qual\ research (QR) \citep{Kelle1997,Lejeune2011,Wiedemann2013,Yuetal2011}.  TM is a computational, \quant\ and big-data method \citep{GandomiHaider2015,GrimmerStewart2013,Wiedemann2013}, whereas QDA is a labour-intensive method due to under-deployment of CAQDAS to some extent \citep{Atkissonetal2016,Kelle1997}.  TM is based on {\it `distant reading'}; i.e., literature is comprehended through aggregating and analysing massive amounts of text rather than studying particular texts \citep{Wiedemann2013,GrimmerStewart2013}.  QDA emphasises {\it `close reading'} of text repeatedly, categorising, interpreting and writing \citep{Wiedemann2013}.  This emphasis restricts QDA to be a time-consuming process, perhaps not a good fit for analysing big data.  No single TM algorithm can fit all research interests \citep{GrimmerStewart2013}.  Similarly, QDA is precisely a collective term for \qual\ analysis approaches, but a general inductive approach is commonly used in health and social sciences \citep{Thomas2006}.  TM involves separating textual data into lexical units to allow indexing and enumerating them for further \quant\ analyses \citep{GandomiHaider2015,GrimmerStewart2013,Wiedemann2013}.  QDA involves  assigning codes to different snippets of text in order to categorise them \citep{Kelle1997}.  Counting is a strategy to derive theme, concept or theory through human interpretation of the data \citep{Thomas2006}.  A widely adapted TM strategy is to customarily discard unimportant words \citep{GrimmerStewart2013}.  Similarly, a recommended QDA strategy is to assign less than 50\% of text to a category to filter relevant information \citep{Thomas2006}.  Both TM and QDA discover knowledge through filtering and categorising text, but employ different techniques.  TM is labelled as an automated text analysis method.  Human input is indeed needed for TM at various stages, e.g., validating/interpreting results \citep{GrimmerStewart2013}.  Some degree of numeracy skill is required to perform QDA.  This paper reports an experiment of integrative applications of TM and QDA techniques with a small-N analysis problem where the text was transcripts gathered from 20 physicians.  It shows that choosing analysis techniques to accord with research philosophies and design involves compromising the conventions of TM, QDA and QR to some extent.  However, various measures show that the semi-automated QDA algorithm retrieved more contextually consistent and relevant information from text than performing TM alone and in a more efficient, systematic and transparent manner than performing QDA alone.

\sec{Methods}
\subsec{The data}
The data were originally collected to develop an ethnographic decision tree model (EDTM) \citep{Gladwin1989} about making clinical decisions for influenza-like illnesses (ILI) during and after the A(H1N1)pdm2009 influenza pandemic in India.  Semi-structured interviews were conducted with 20 purposively chosen practitioners of Allopathy, Homeopathy or Ayurveda.  The sample size is determined by data saturation -- a widely adapted criterion for determining sample size for qualitative research \citep{Mason2010}.  The interviews were guided by an instrument (with 22 open-ended questions) and a clinical vignette (with 14 open-ended questions), and audio-taped and transcribed/translated into English text.  The instrument was applied to collect training data from 10 participants to build a preliminary model.  The vignette was then designed and used together with the instrument to collect testing data from another 10 participants to validate the model.  Further details are available from \citet*{Ahankarietal2017}.

\subsec{The analysis procedures}
The analyses were performed by {\tt R} version \mbox{3.2.3} \citep{Rcore2015} and various packages including {\tt tm} version \mbox{3.2.3} \citep{Feinereretal2008}, {\tt RQDA} version \mbox{0.2-8} \citep{Huang2016} and {\tt igraph} version \mbox{1.0.1} \citep{CsardiNepusz2006}.  Word clouds were depicted by WordCloud version \mbox{1.6.0} \citep{Mueller2019} in Python version \mbox{3.6} \citep{VanRossumDrake2009}.  All analyses were executed by a Ubuntu \mbox{12.04.5} LTS server with 10GB RAM.  The CPU speed was \mbox{2.40GHz}.  The interview transcripts were converted from pdf to html format by (Apache) Tika (\url{https://tika.apache.org/1.3/formats.html}) before being imported into the {\tt R} environment.

The TM process was based on two assumptions: (i) documents are a `bag of words' where word order will not change the nature of the sentence, and (ii) a simple list of unigrams (individual or hyphenated  words) is sufficient to convey the general meaning (or context) of a text \citep{GrimmerStewart2013}.  The baseline procedure was as follows:

\begin{enumerate}
\setlength{\itemsep}{2pt}%
\setlength{\parskip}{1pt}%
  \item Cleaning the raw data by performing spell-check to eliminate typos;
  \item Reducing dimensionality of the text by converting words to base forms (e.g., `come' would replace `comes', `came' and `coming');
  \item Replacing `contractions' with longhand words enlisted by the {\tt R qdapDictionaries} package \citep{Rinker2013};
  \item Splitting the transcripts into unigrams (individual or hyphenated words) based on a manually validated vocabulary;
  \item Eliminating domain-specific stopwords (very common words) gathered from PubMed or EBSCOhost;
  \item Indexing each unigram to obtain a lexicon that enumerates the respective occurrence(s) in each transcript; and
  \item Identifying important unigrams by discarding unigrams with occurrences less than 1\% or greater than 99\% in the respective transcript sets.
  \end{enumerate}

\noindent Point~7 was a deviated approach suggested by \citet{HopkinsKing2010}.  If only unigrams with the highest frequencies were retained, then the opinions of minority participants would probably be entirely discarded.  It contrasts with the intent of applying purposive sampling that aimed to  deliberately include outliers in the research sample and let the exception prove the rule (\cf\ \citet{Barbour2001}).  After the pre-processing stage (points \mbox{1--3),} a vocabulary of all the transcripts was extracted by using a language processing tool, {\tt treetagger} (\url{http://www.cis.uni-muenchen.de/~schmid/tools/TreeTagger/}).  The analyst manually validated the vocabulary with references to the linguistic structure information provided by {\tt treetagger}.  The validation process was also assisted by programmes written by the analyst in Visual Basic Applications for spreadsheets (\url{https://en.wikipedia.org/wiki/Visual_Basic_for_Applications}) to highlight the words whose base form was the same.  The results helped the analyst manually re-categorise words to reduce the text dimensions based on her domain-specific knowledge to some extent.  For example, `patiently' was replaced by `patience' to distinguish it from `patient' (a person) (see Table~\ref{table-1}\hide{1}).  The vocabulary served as a dictionary (i) to group original words with the same base form (or unigram/root) to reduce data dimensionality at the TM stage, and (ii) to find out the original words to be coded at the QDA stage.  That is, lemmatisation was applied to define the unigrams based on a dictionary \citep{GrimmerStewart2013}.  Table~\ref{table-1}\hide{1} presents an excerpt of the vocabulary and explains how it was built and applied.  

\begin{table}[!b]
 \centering
 \includegraphics[trim=290 1380 290 280,clip,scale=0.38]{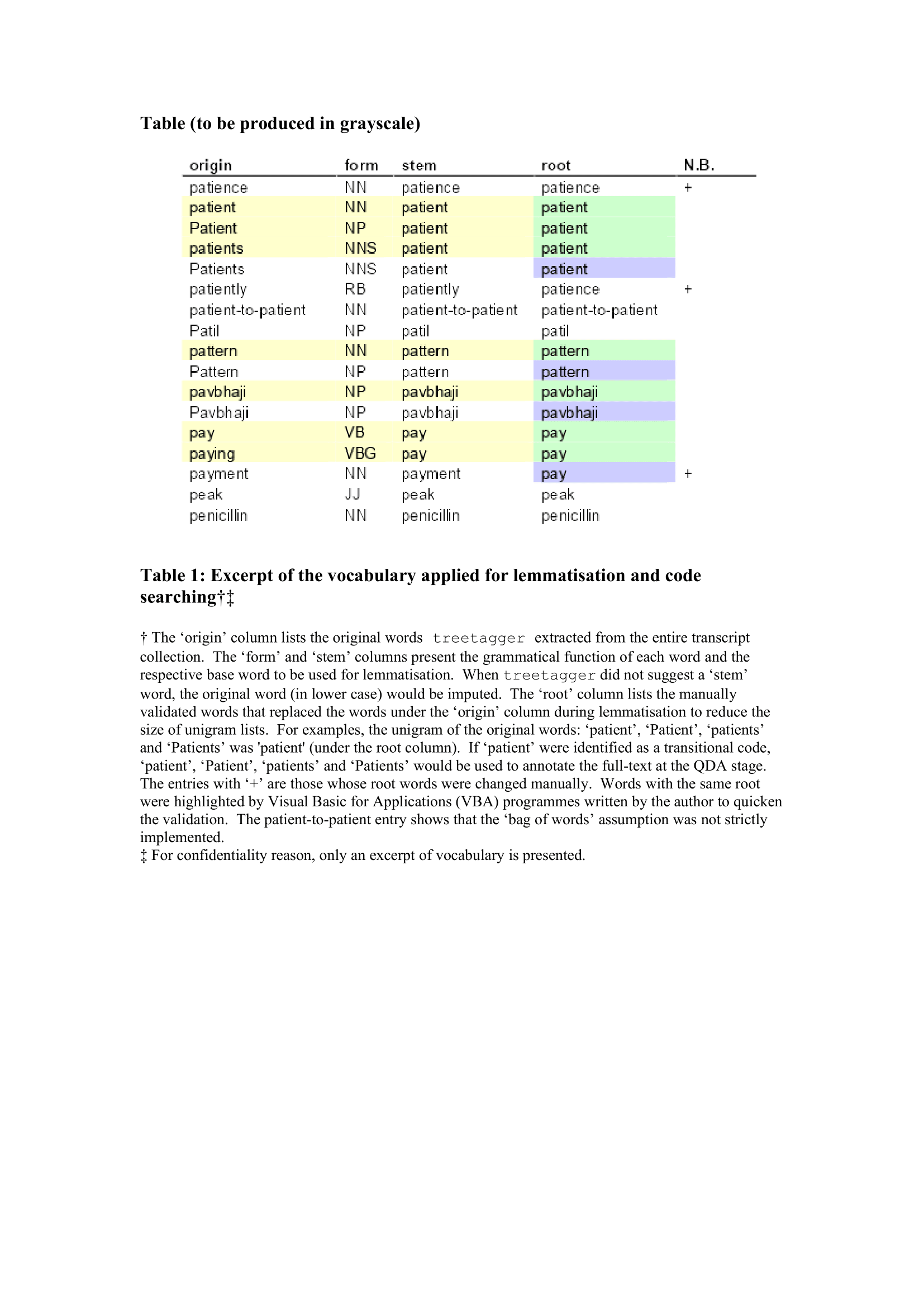}
 \caption{Excerpt of the vocabulary applied for lemmatisation and code searching$\dagger$$\ddagger$}
 \label{table-1}
  \includegraphics[trim=250 780 250 1090,clip,scale=0.36]{article-TM-Table-1}
\end{table}

\begin{table}[!ht]
  \includegraphics[trim=250 520 250 190,clip,scale=0.36]{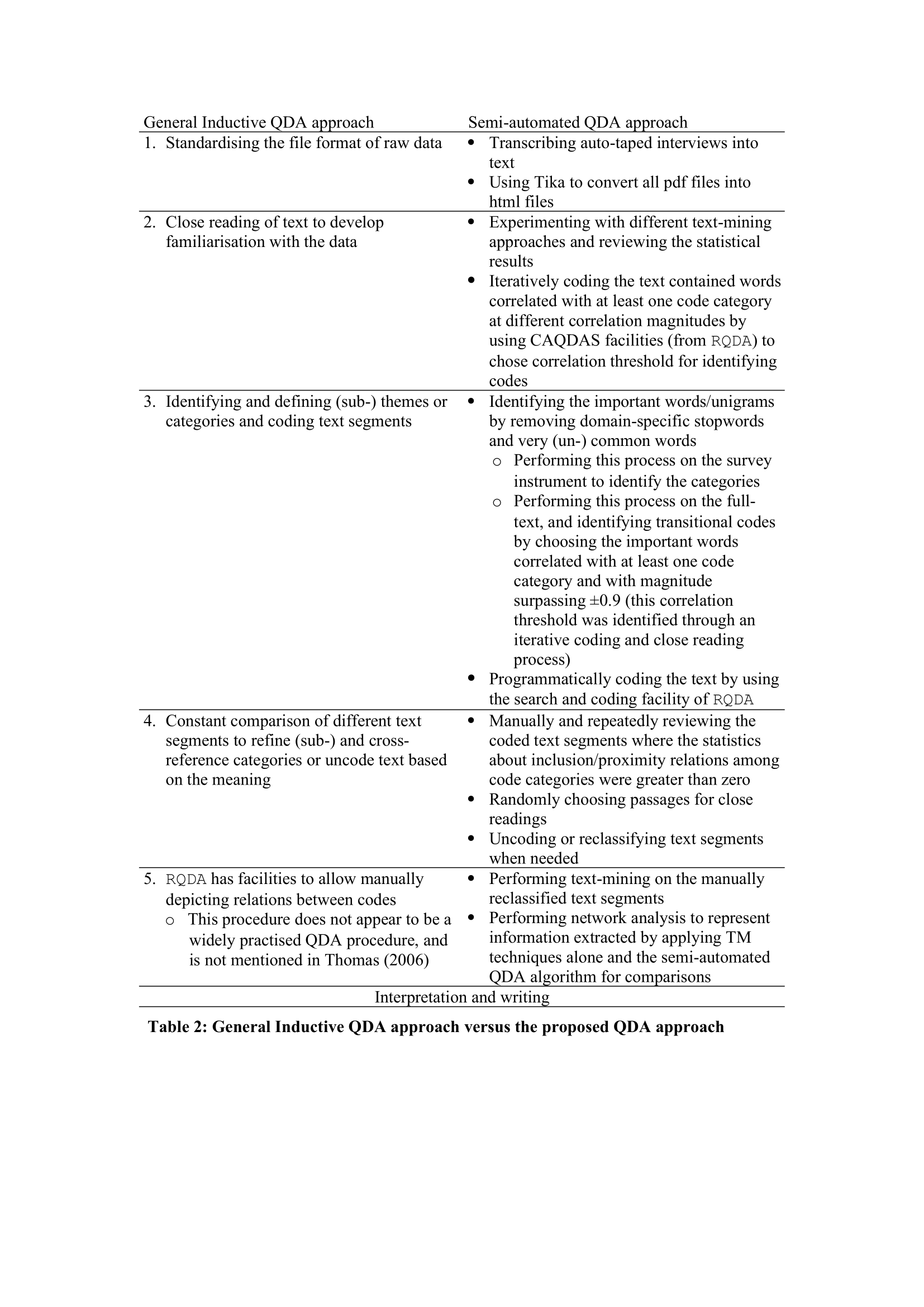}
  \caption{General Inductive QDA approach versus the proposed QDA approach}
  \label{table-2}
\end{table}

TM was performed on (a) the 18 open-ended questions used to gather training data (the four questions regarding participant profile were excluded), then the 32 open-ended questions used to gather testing data, and (b) the training and testing transcripts.  This procedure results in four sets of unigrams and occurrence statistics.  Stage (a) was performed to form bases to operationalise the idea that interview questions are a good basis for identifying key themes from qualitative data \citep{Rowley2012}.  The important unigrams gathered at stage (a) should reflect expert opinions, and be reliable (code) categories to classify coded text at the QDA stage.  A transitional code is an important unigram identified at stage (b) and significantly correlated with at least one category of its corresponding text set.  It could fall into more than one category.  A code is an original word enlisted by a transitional code.  The correlation threshold $\pm$0.9 was chosen through coding the transcripts iteratively by using the automated search and coding facilities of {\tt RQDA}.  The search started by identifying codes belonging to transitional codes with correlation $\pm$1.  The entire sentence containing a code would be annotated.  It was followed by using the next 0.05 highest correlation magnitude ($\pm$0.95) to identify codes and annotate the transcripts.  The search process was halted when only about 50\% of the transcripts were coded (albeit, only less than 50\% of the raw text needs to be coded and categorised \citep{Thomas2006}).  In the same rubric, a category correlated with an excessively long list of codes might not extract crucial information.  For example, the category `common' extracted from the questions used to collect training data contained 28 transitional codes, but the second longest list contained 12.  So, only the codes which fell in the `common' categories were annotated.  This approach might avoid substantially eliminating outlier views (\cf\ \citet{Barbour2001}).  The objective of the study was to derive an EDTM model where each decision is represented as a dichotomous outcome \citep{Gladwin1989}; e.g., prescribing antibiotics or not.  Coding paragraphs with `yes/Yes' or `no/No' was a consistent technique.  Widely used words for conducting content analysis were also coded -- the words indicated negation, amplification, deamplification, positive or negative, available from {\tt qdapDictionaries}.  The statistics about (inclusion/proximity) relations among code categories were applied to identify text segments needed for close reading (see \citet{Huang2016}, for details about relations statistics).  It sometimes involved close reading of passages/text segments near the automatically annotated text, so as to recode the text.  Some passages were randomly chosen for close reading to check the logic flow of the coded text.  Domain-specific knowledge was the basis for reclassification and uncoding of text.  When no clear-cut coding decision could be made, recoding/reclassification was guided by the frequencies of 'yes/Yes', 'no/No' and the words used to perform content analysis.  This technique originated from automated content analysis \citep{GrimmerStewart2013}, but was adapted to guide recoding/reclassification.  TM was performed on the manually recoded/validated text segments to enable comparing the results obtained by performing TM alone on the full-text.  Table~\ref{table-2}\hide{2} summarises the techniques employed to accomplish the recommended procedure to implement a widely adapted general inductive QDA approach \citep{Thomas2006}.

Network analyses were performed to depict and validate the connections between the unigrams extracted from the transcripts and the manually recoded text segments.  The induced subgraphs were derived from the scaled matrices of unigram occurrences obtained from TM.  Occurrence statistics were rescaled into `1' (`0') to represent the presence (absence) of each unigram in (from) each text unit.  Modularity clustering was performed on each induced subgraph.  The clustering algorithm involved iteratively removing the edge with the highest edge-betweenness score until the vertices became segregated into clusters \citep{NewmanGirvan2004}, each consisting of at least two vertices.

\sec{Results}
The transcripts have 52,023 words.  It took \mbox{2--3} days to prepare a manually validated vocabulary to assist performing TM and automated coding, and four days to uncode and reclassify the coded text manually.  The automated text analysis approach retrieved 59 out of the 242 manually reclassified/validated text segments.  So, the coding concordance is about 24.4\%.  However, in almost all instances, the required text segments could be found within the same or \mbox{1--2} nearby paragraph(s).

\begin{figure}[!b]
 \begin{tabular}{lc}
Training data:& \\[-12pt]
 &\includegraphics[scale=0.24]{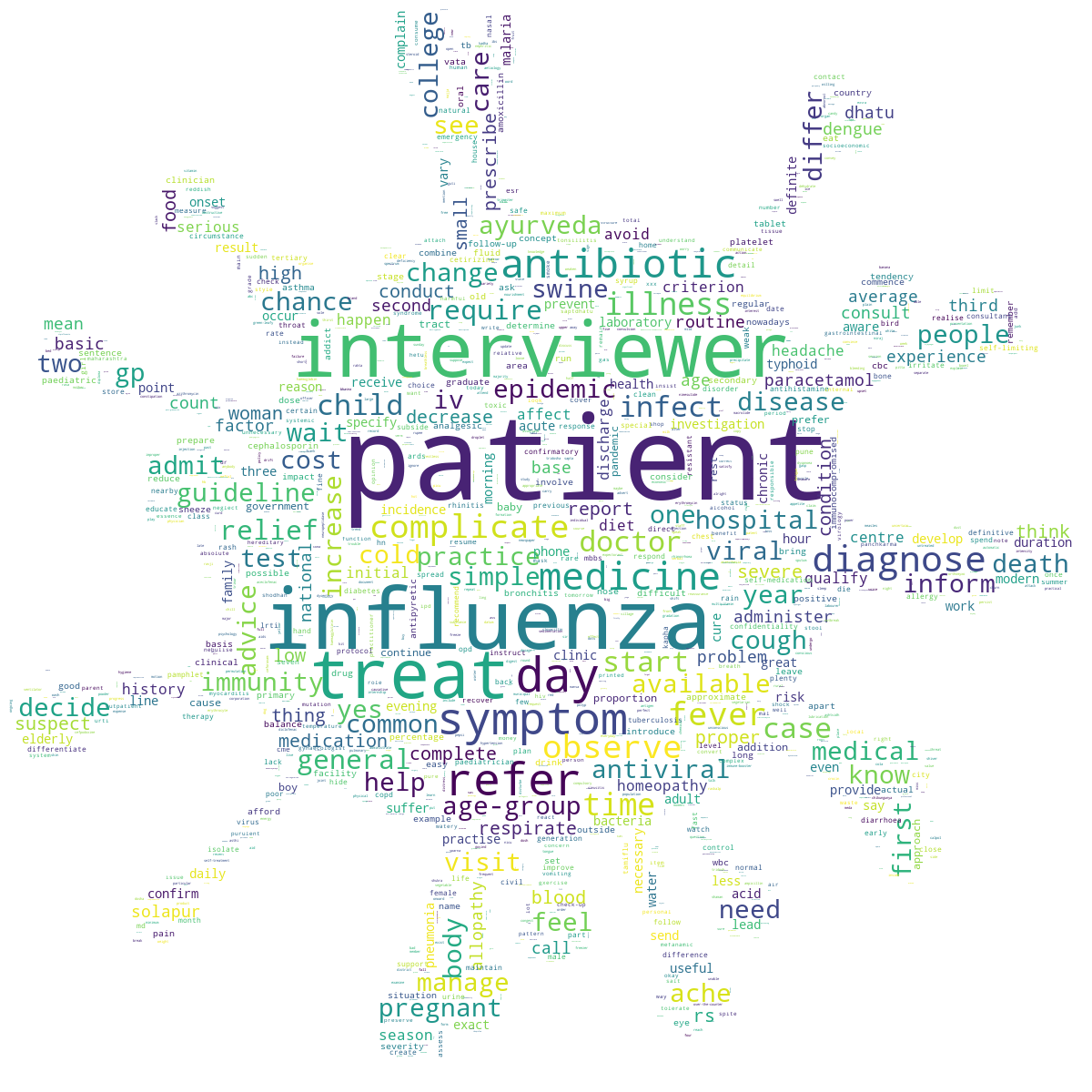}\\
Testing data:&\\[-12pt]
 &\includegraphics[scale=0.24]{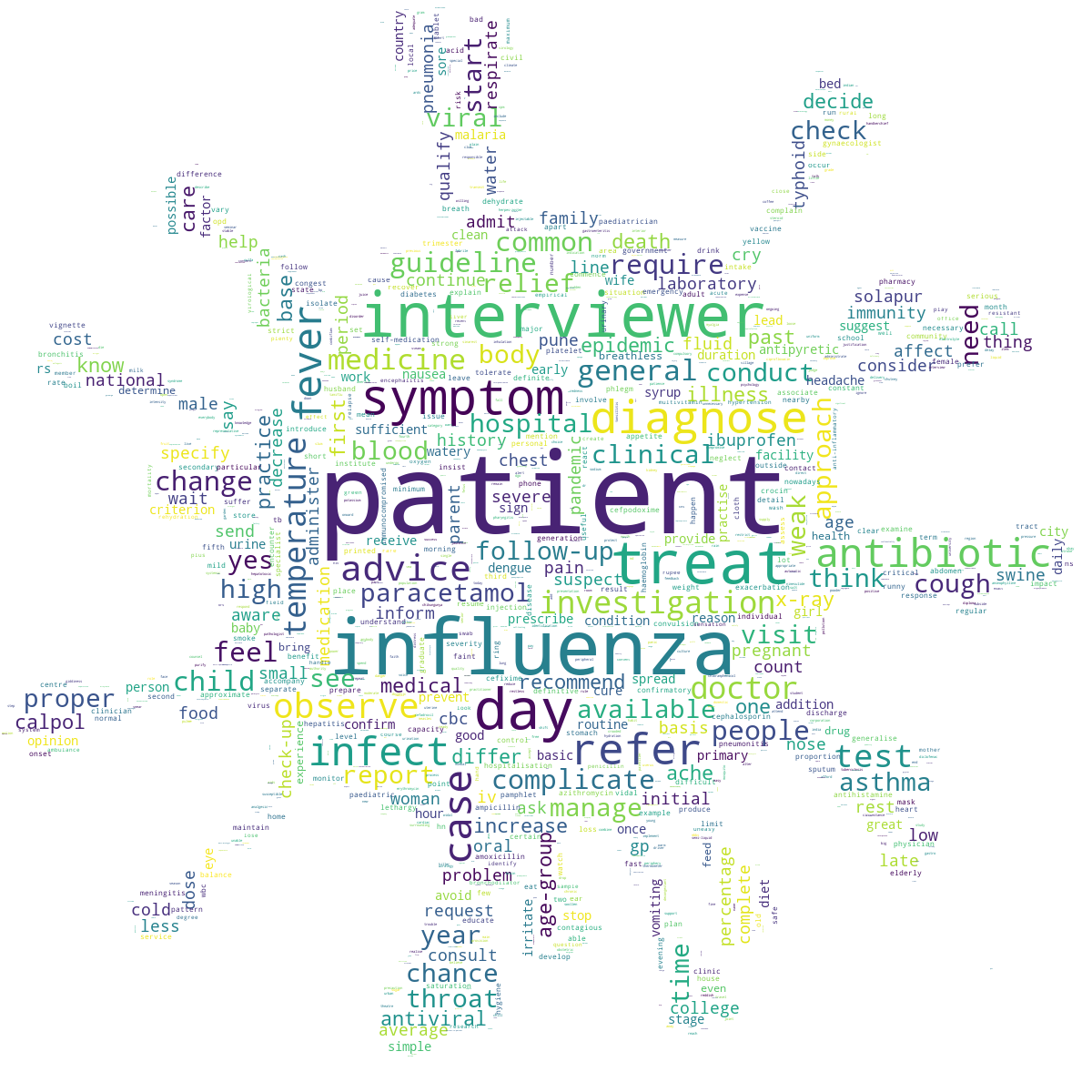}\\[-6pt]
 \end{tabular}
 \caption{Word frequencies of the transcripts after removing domain-specific stopwords}
 \label{figure-1}
\end{figure}

\begin{figure}[H]
  \centering
  \includegraphics[trim=250 320 250 200,clip,scale=0.36]{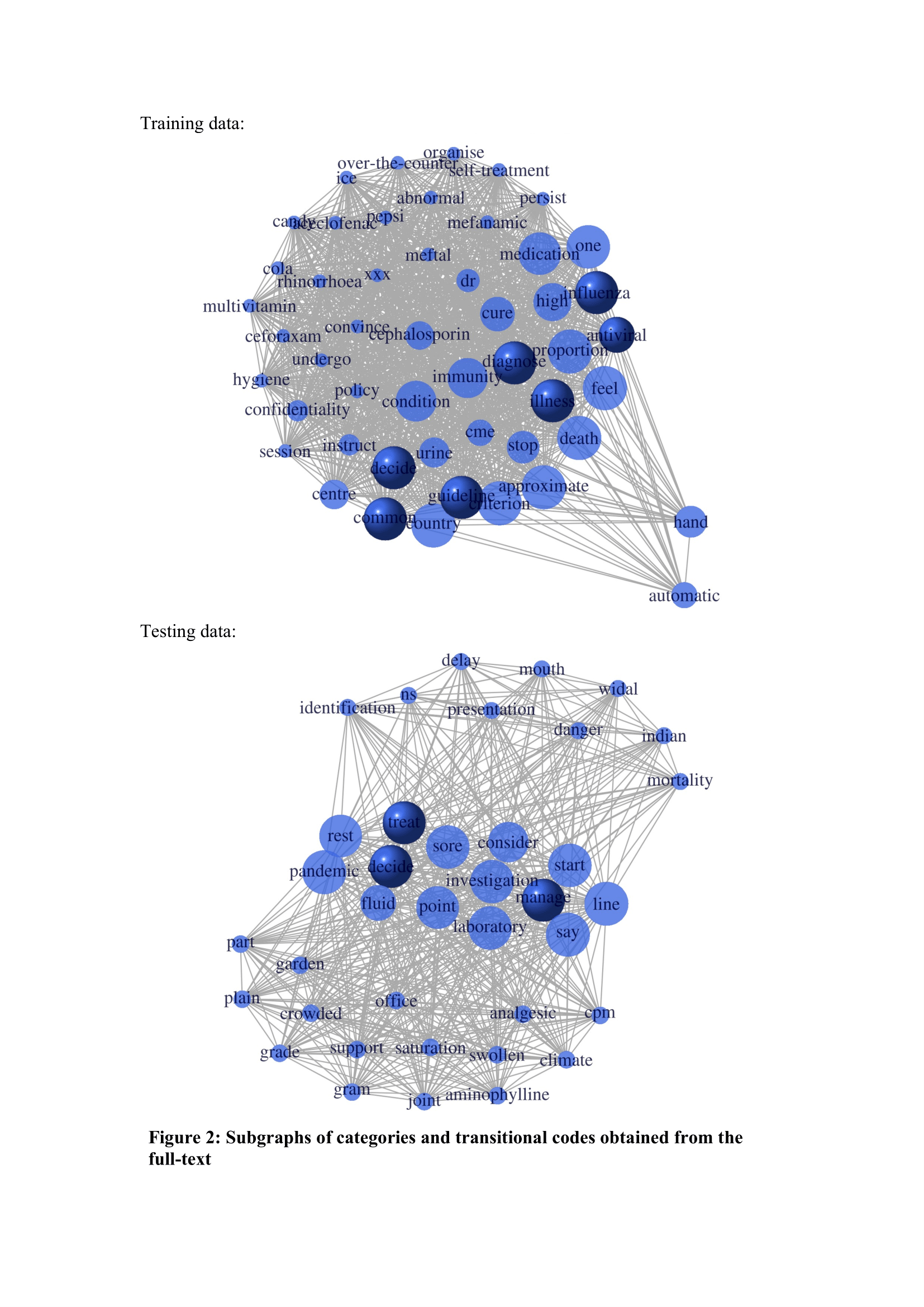}
  \caption{Subgraphs of categories and transitional codes obtained from the full-text}
  \label{figure-2}
\end{figure}

\begin{figure}[H]
  \includegraphics[trim=250 260 250 210,clip,scale=0.35]{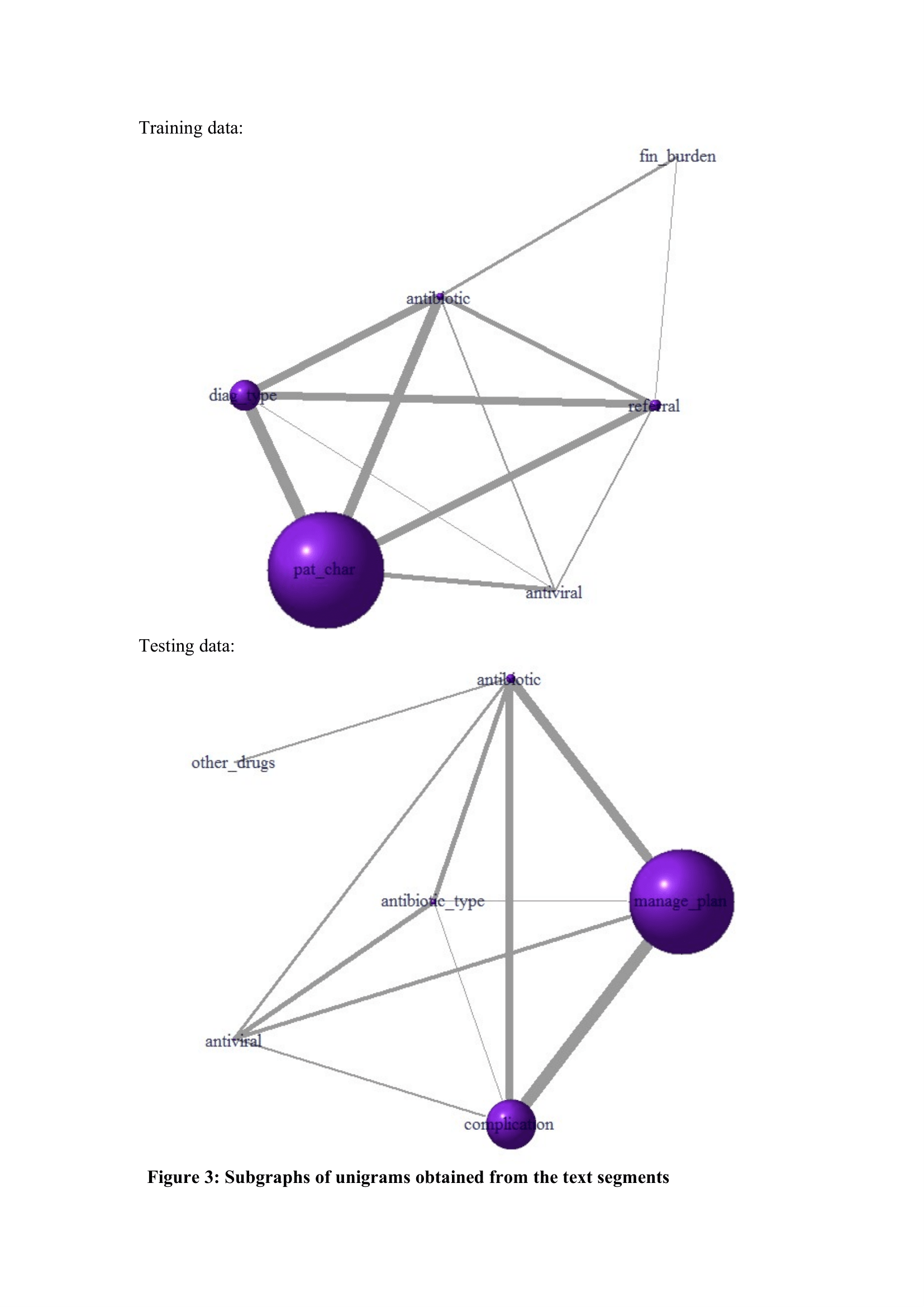}
  \caption{Subgraphs of unigrams obtained from the text segments}
  \label{figure-3}
\end{figure}

Figures~\ref{figure-1}--\ref{figure-3}\hide{1--3} represent results obtained at different stages of the analyses (outlined in Table~\ref{table-2}\hide{2}).  Figure~\ref{figure-1}\hide{1} depicts the frequencies of unigrams obtained from the training and testing transcript collections by performing procedures \mbox{1--2} after removing the domain-specific stopwords.  The size of each unigram represents the relative frequency in the respective transcript set.  The words with highest frequencies seem to overlap substantially.  The data structures are qualitatively similar.  The topic was rightly captured as interview[s] regarding clinical strategies and health management applied to patient[s with] influenza\mbox{[-like]} illness[es].  Figure~\ref{figure-2}\hide{2} presents the subgraphs of the categories or transitional codes (important unigrams surpassing the $\pm0.9$ correlation threshold).  Vertices depicted as spheres (circles) represent the categories (transitional codes).  The size of each vertex represents its degree (the number of connections with other vertices) relative to the maximum degree of the original graph.  Each edge represents a link from one unigram to another one.  The width of each edge represents its weight (distance from one vertex to another) relative to the maximum weight.  The bigger (thicker) a vertex (edge) is, the more important it is.  All categories and transitional codes are connected to different extents.  Due to the complexity of the subgraph, only the sizes of vertices can be compared visually.  The categories appear more important than most transitional codes.  TM identified 68 (82) unigrams from the training (testing) data.  Network analyses suggested forming one module with 50 (38) unigrams from the training (testing) data.  3 (5) out of 10 (8) categories of the training (testing) data were eliminated.  The $p$-value of the Wilcoxon statistic of the training (testing) data is $p$=0.000 ($p$=0.046).  The module of the training data is of a better fit, because its $R^2$ statistic of degree distributions (0.777) is higher than that of the testing data (0.595).

\newpage
\newpage

\newpage
\newpage
\begin{table}[H]
  \includegraphics[trim=250 310 250 190,clip,scale=0.35]{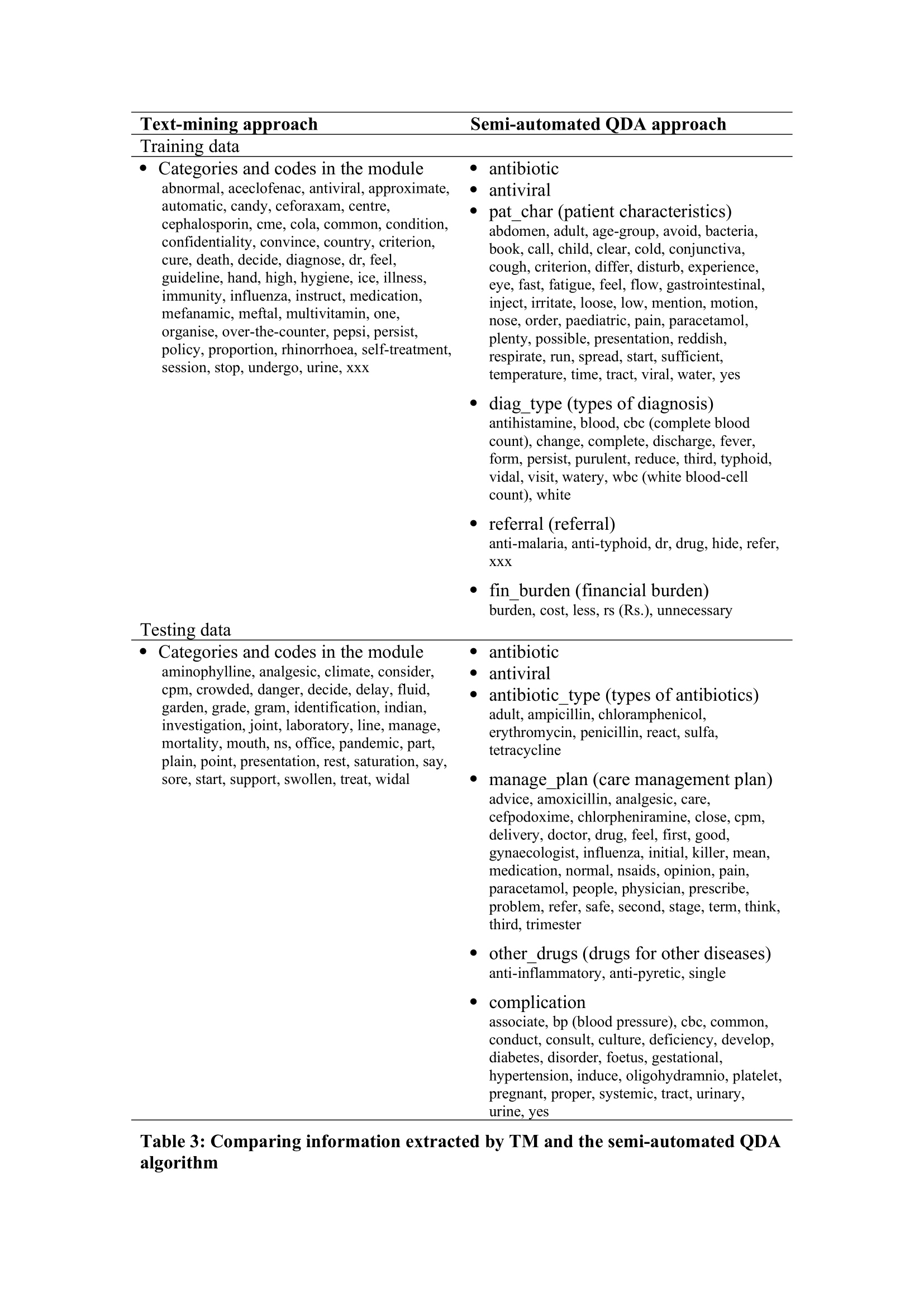}
  \caption{Comparing information extracted by TM and the semi-automated QDA algorithm}
  \label{table-3}
\end{table}

Figure~\ref{figure-3}\hide{3} depicts the induced subgraphs of the unigrams extracted from the manually recoded/validated text.  It represents the results obtained by performing QDA guided TM results.  Modularity clustering was performed to find out the immediate neighbour vertices of `antibiotic'.  `antiviral' was then included to depict its connections with those vertices and `antibiotic'.  The `antibiotic' and `antiviral' spheres are isolated vertices.  The other spheres represent modules of vertices.  The size of each sphere represents the aggregated degree of each module/vertex relative to the total degree of the respective subgraph.  The width of each edge represents the relative aggregated weight.  For the training data, the four \mbox{(anti-)} modules are pat\_char ($p$=0.000), diag\_type ($p$=0.000), referral ($p$=0.001) and fin\_burden ($p$=0.424) (see Table~\ref{table-3}\hide{3} for the components of each module).  The fin\_burden module is an anti-module, because its Wilcoxon statistic is non-significant $p$=0.424 (>0.05).  Its member vertices have more connections with vertices from other modules than vertices within fin\_burden.  For the testing data, the four \mbox{(anti-)} modules are antibiotic\_type ($p$=0.000), manage\_plan ($p$=0.000), other\_drugs ($p$=0.047) and complication ($p$=0.114).  In both cases, `antibiotic' connects with `antiviral', and not all neighbour vertices of `antibiotic' connect with `antiviral'.  The edge widths and the sizes of the vertices indicate that the neighbour vertices of `antibiotic' form stronger connections with `antibiotic' than `antiviral'.  `antibiotic' is more important than `antiviral'.

Table~\ref{table-3}\hide{3} lists the unigrams that formed Figures~\ref{figure-2} and \ref{figure-3}\hide{2 and 3}.  The first (second) column shows the results obtained from the full-text (text segments).  It shows that the full-text and text segments are contextually consistent.  For example, the module obtained from the training full-text reveals various antibiotic names and terms associated with different diagnoses and referral [to xxx hospital/centre or Dr. xxx].  The modules of the testing full-text contain terms associated with a wider range of topics including [antibiotic] lines, [clinical] presentation, pandemic, laboratory [tests], treatment, etc.  The vertices of antibiotics appear more often than other medications.  The connectivity of `antibiotic' with other unigrams appears to be an important construct.  The text segments of the testing data inherently capture a wider range of topics than those obtained from the training data (the full-text of the testing data actually contained more important unigrams than the training data).  The manage\_plan module is about medication, clinical presentation, referral [to an allopathic physician/doctor or a gynaecologist], laboratory diagnosis, etc.  These concepts can be seen from the diag\_type module extracted from the training data.  The diag\_type module reveals diagnoses applied to check the severity [of illness], pathology and possible complications.  The subgraphs of the text segments appear to capture information more specifically than those of the full-text.

\sec{Discussion}
This paper presents an example of applying freely available computational facilities to enable a novice analyst of TM and QDA to code, validate and exact information from 20 sets of transcripts in \mbox{6--7} days (and in a presentation format), whereas the comparable tasks took an experienced \qual\ analyst months by using a conventional QDA approach.  Network analysis shows that both the TM and semi-automated QDA algorithms extracted contextually consistent and relevant information.  Employing the semi-automated QDA algorithm is preferred, as it extracted more specific information than applying TM alone and turned QDA into a more efficient, systematic and transparent process.

The transcripts consisted of 52,023 words.  The size is approximately \mbox{11--18} articles published in top-tier peer-reviewed journals.  TM identified 68 (82) important words/unigrams from the training (testing) full-text as categories and transitional codes.  Network analysis suggested retaining 88 unigrams (see Figure~\ref{figure-2}\hide{2}).  The testing text might have captured more diverse topics than the training text, as it contained more important unigrams and the module was formed by fewer unigrams.  The vignette seems to serve as a function to unveil more hypotheses.  TM appears efficient in filtering relevant information.  A \qual\ analyst may deny that a handful of words can represent the complexity of \qual\ data.  Indeed, TM only retrieved 24.4\% of the manually recoded text segments.  This concordance of coding appears poor by the recommended standard of inter-rater reliability rate (IRR) \citep{Cicchetti1994}.  However, this measure inappropriately downgraded the performance of TM, because the relevant text segments could mostly be found within \mbox{1--2} nearby paragraph(s) of the automatically coded text.  It is hard to \mbox{(dis-)} prove this claim, and also reaching acceptable IRRs.  We can inspect the procedures for reaching agreements among coders, but not the objectivity of the decisions.  However, the claim is indirectly backed by the fact that the TM results accomplished the \mbox{(re-)} coding and extracting of validated data in a presentable format in \mbox{3--4} days.  At an earlier research phrase, it took an experienced \qual\ analyst months to obtain the validated information in verbatim form (a widely used QDA reporting format).  The major objective of the study is to understand the clinical strategies physicians used to diagnose and manage influenza-like illnesses during and after the A(H1N1)pdm2009 influenza pandemic.  Figures~\ref{figure-2} and \ref{figure-3}\hide{2 and 3} shows that both the information exacted by TM alone (from the full-text) and information manually validated through a QDA process present antibiotic and antiviral as a theme.   The extracted information overlaps with certain sub-themes extracted by another analyst by performing thematic analysis (see also Figure~\ref{figure-a1}\hide{A1} in the appendix).  The subgraphs show that the (semi-automated) QDA algorithm did exact more specific information than TM alone.  A mainstream statistician may reject judging the consistency, relevance and quality of information extraction by comparing the meaning of subgraphs.  Subgraphs are a kind of cluster analysis results.  The interpretation can be subjective.  S/he may propose performing a Steiger $z$-test \citep{Steiger1980} to test the differences in the degrees of correlations of unigrams obtained from the training and testing text sets.  This test is reliable if a domain-specific dictionary is available to assist precisely grouping words of the same roots together.  If such a dictionary did exist, Natural Language Processing (NLP) would not only be successfully applied to discover knowledge from electronic health records (EHRs) in certain domain-specific systems \citep{Feldmanetal2016}.  The semi-automated QDA algorithm inherits the convention of QR, which allows researchers' (and/or the participants') subjective experience to be included in the analysis process \citep{Eversetal2011,GreenhalghTaylor1997,GreenBritten1998,Kelle1997,NobleSmith2015,SuttonAustin2015}.  The appraisal criteria established for evaluating \quant\ research cannot be applied to QR without justification \citep{GreenhalghTaylor1997,GreenBritten1998,NobleSmith2015}.

Implementing the proposed TM and QDA algorithms involves applications of computational and statistical skills, but not to a sophisticated level.  Certain CAQDAS products are packed with the facilities used in this experiment, and come with a graphical interface \citep{Lejeune2011,Wiedemann2013}.  Advanced technologies are not needed either.  The analyses were conducted by a computer with low RAM and speed by average standards.  A more powerful computer will only further scale up the process slightly, as it involved human input repeatedly at various stages.  For example, validating the vocabulary used for lemmatisation and close-reading of text segments were meant to be manual procedures.  The TM and QDA algorithms were built on widely used statistics.  Frequency and/or Pearson correlation were used to identify the (code) categories and transitional codes from the lexicons of unigrams.  Like correlation, the (inclusion/proximity) relation statistics of categories used to guide reviewing the automatically coded text are also association measures.  Network analysis is an optional procedure, although it has a role in validating and further filtering information.  It is not an interim analysis between TM and QDA.  The interpretation of the text was not influenced by the network analysis results.  \Qual\ researchers may see using network analysis to present QDA results as controversial.  It is a widely applicable method for presenting information.  For example, it is a compulsory procedure of meta-analysis \citep{Huttonetal2015}.  This approach is more efficient than manually depicting the relations of codes by using the facility that comes with some CAQDAS.  Researchers can also report the QDA results in verbatim form.  \Quant\ researchers may argue that the analysis procedures are not scientifically rigorous given that the correlation threshold (0.9) used to search for codes was identified through iteratively \mbox{(re-)} coding the text until a judgemental principle was fulfilled (less than 50\% of the text needs to be coded and categorised \citep{Thomas2006}).  In fact, statistics is an art and science of learning from data \citep{AgrestiChristine2016}.  The practice always involves subjective elements and changes to accord with specific occasions.  If evidence must be in numerical form in order to claim to be scientific, then the analyses were substantially guided by `numbers'.  {\tt treetagger} used to build a dictionary for lemmatisation and other dictionaries employed to assist recoding were the results of dedicated research efforts of various domains over decades \citep{Rinker2013}.  Their role is similar to establishing external validity by using reliable external data.  Validating coding did not purely rely on applying domain-specific knowledge of one single analyst.  The constant comparison process was complemented by reviewing randomly selected passages to repack the coding.  This technique is akin to applying stepwise variable elimination to derive parsimonious regression models.  Obtaining approximation from randomly chosen subsamples is a widely used statistical technique; it is the core of the widely applied bootstrap technique \citep{EfronTibshirani1994}.  The analytic procedures are probably more rigorous, transparent and systematic than many mainstream analytic methods and the conventional QDA methods.  The approach should be easy to understand and implement.

The algorithms incorporated expert opinions of various domains into the design, as analytic techniques were carefully chosen to accord with the research philosophies and study designs \citep{GreenhalghTaylor1997,GreenBritten1998, Grimmer2015, Kelle1997, NobleSmith2015,Thomas2006,Wiedemann2013}.  For example, the data were collected and analysed by training and testing sets to accord with EDTM approach \citep{Gladwin1989}.  Ethnographic decision modelling is known for being able to achieve at least 80\% predictability of behaviours under studies \citep{RyanBernard2006}.  The study design may contribute to being able to extract consistent and relevant information from the two transcript sets.  TM is a big-data/\quant\ method \citep{GandomiHaider2015,GrimmerStewart2013,Wiedemann2013}.  The capacities of TM will be undermined when TM is applied to small-N data, because not all quantitative techniques are capable in obtaining robust statistical results from small-N data.  In this experiment, the analyses were guided by statistics whose robustness is neither dependent of sample size (i.e., frequency) nor undermined significantly when sample size is small (i.e., association measures).  If the data were gathered from fewer participants, Fisher's exact test of independence might be performed to validate the associations of unigrams \citep{Raysonetal2004}.  Automating QDA does not necessarily turn QDA into another big-data or \quant\ method.  Sense-making and domain-specific knowledge are needed to interpret the results and modify the vocabulary used for lemmatisation.  The lemmatisation approach permitted customarily splitting text into different sizes based on expert opinions to some extent.  It is different from the widely applied tokenisation approach where text is uniformly split into one single size.  How the words are split is considered important for preserving the context of text on occasions \citep{GrimmerStewart2013}.  It relaxes the `bag of words' assumption, which appears unpalatable to \qual\ researchers.  This QDA process is definitely not mechanical.  TM can be adapted into QDA without changing the QDA process of creativity and opportunity for serendipity into a mechanical process of performing some code plans on large document collections \citep{Wiedemann2013}.

\Qual\ researchers often claim that they `allow the theory to emerge from the data' \citep{StraussCorbin1990}.  For outsiders, this claim is mysterious \citep{Thorne2000}.  The literature is difficult to comprehend or use \citep{Thomas2006}.  The proposed QDA algorithm is not only efficient, but also a systematic and transparent process.  It will help in making QDA an easier approach to understand and implement.  It did enable a novice of TM and QDA to accomplish the tasks quickly.  As yet, an application of CAQDAS is sometimes mistakenly referred to as an analytic method \citep{Eversetal2011} (see \citet{Lunnyetal2016}, for an example).  The proposed QDA approach allows presenting the words used to code the text or classify the text segments, and the relation statistics used to guide constant comparisons and recoding.  External data was also used to guide code repacking.  It is a systematic approach, but not a data driven approach.  The analyst was the judge of the final results at various stages.  Conventionally, independent multiple coding guided by IRR is a popular technique to establish the trustworthiness of QDA \citep{Barbour2001,GreenhalghTaylor1997}.  This practice is either too costly or not always a viable option \citep{Barbour2001}.  A cost-effective alternative was employed.  Initial coding was automatically done based on TM results.  Subsequently, relation statistics were obtained to guide repacking the coding.  The TM results together with relation statistics served the function of a codebook.  This approach will remain an efficient method even if the text collection is huge.  Developing a codebook is a time-consuming process, and feasible only when the text collection is small \citep{Mcalisteretal2017}.  In that study, with the help of a codebook, the IRR of initial coding could be as low as 40\% (poor agreement) \citep{Mcalisteretal2017}.  The fallible measure of concordance of TM results with that of QDA only indicates that the TM algorithm is not a standalone algorithm by QDA standard.  However, it is not a basis for dismissing the application of TM as an assistive QDA technologies.  The entire analysis process was completed in \mbox{6--7} days and extracted relevant and consistent information in a presentable format.  If CAQDAS facilities were merely used to manage data, it could take hours or even days to interpret and compare several retrieved text segments \citep{Kelle1997}.  Interpreting the text properly involved domain-specific knowledge.  It is unlikely that a reliable TM algorithm can be developed in one single experiment.  Also, the literature provides little information about how IRRs coders can obtain at an initial coding stage.  How coders reach agreements is actually not a well articulated topic, and IRR is not a uniquely defined measure \citep{Mcalisteretal2017}.  Indirect evidence suggests that the TM algorithm might not perform noticeably worse than humans.  An experiment showed that the six experienced coders, who analysed the same text collection, packaged coding frameworks in considerably different ways \citep{Armstrongetal1997}.  An experiment with five CAQDAS products confirmed that `[t]here is no fixed way of interpreting qualitative data' \citep{Eversetal2011}.  QDA results are sometimes seen as an interpretive framework `converged' by researchers of the study \citep{Moretetal2007}.  This criticism does not apply to the proposed QDA algorithm, although it still presents subjective elements like vastly many analytical algorithms do.  This experiment replicates a conclusion of certain experiments where TM could assist scaling up manual curation, but was not a substitute for it since humans were more reliable in extracting facts from text \citep{Winnenburgetal2008}.

The analysis techniques were chosen to implement a QDA general inductive approach \citep{Thomas2006}.  It is not a suggestion for applying one QDA approach to all types of unstructured text.  Instead, it is an attempt to develop a widely applicable semi-automated QDA algorithm.  The TM procedures deviate from a completely automated approach.  It allowed the analyst to revise the dictionary used to perform lemmatisation, choose the criterion for identifying codes for initiate coding and act as the final judge of coding.  It is different from a supervised machine learning approach that where text is manually and iteratively \mbox{(re-)} coded subject to what the machine learns from the manual coding until a given level of coding concordance is reached \citep{Atkissonetal2016,GrimmerStewart2013}.  Further experiments are needed to examine the applicability of the QDA algorithm to other research topics and data of different structures and sizes.  If a domain-specific medical dictionary existed, lemmatisation would probably be further improved.  Alternatively, lemmatisation may be carried out based on a vocabulary revised by an expert panel.  For a text collection of enormous size, the panel may review the vocabulary taken from several randomly chosen subsamples.  Another alternative is to build a dictionary of keywords extracted from articles shortlisted through a systematic literature search process.  The required facility is freely accessible (see \url{https://elizagrames.github.io/litsearchr/#/about}, for example).
One may consider the experiment redundant, since it is within human capacity to analyse the text.  The data were small-N data and hardly fitted into the oft-cited \mbox{3-V} definition of big data \citep{GandomiHaider2015}.  It is certainly within the capacity of a human to analyse them.  However, investigating this analysis problem is, at least, critical to improving health-related sciences.  Medical practitioners need current best qualitative evidence to deliver medicine tailored to serve each patient as a unique individual \citep{Tonelli2018}.  Personalised medicine is likely to be evolved into standard practice.  QR is growing rapidly, but remains a small body of the medical literature \citep{Shuvaletal2011}. QR usually gathers data from a handful of research subjects \citep{Kelle1997,Lejeune2011,Wiedemann2013}.  They are small-N data, but can also be big data.  Conducting QR usually involves analysing vast amounts of textual data originated or converted from various sources and/or formats \citep{Popeetal2000,SuttonAustin2015}.  Without the help of TM, delivering timely \qual\ evidence to support approved medical practice is impossible.  In particular, medical knowledge was estimated to double every seven years in 2010 and will double in 73 days by 2020 \citep{Densen2011}.  This experiment is a preparation for mitigating this issue.  The discussion also shreds light on the development algorithms to deploy EHRs for better care.  Remarkable progress has been achieved by using NLP to discover knowledge from those systems \citep{Feldmanetal2016}.  Due to the complexity of medicine and the language, no single algorithm is universally applicable to reliably extract the information from systems of different domains \citep{Feldmanetal2016}.  Improving artificial intelligence in line with QR/QDA principles could, at least, be an interim solution.

This paper explores the challenges of applying a semi-automated QDA approach to analysing small-N unstructured textual data.  It involves compromising the conventions of QR, QDA and TM to some extent.  However, it shows promises in reliably discovering knowledge from text in a more efficient, systematic and transparent manner than relying on human interpretation.  This paper also paraphrases the techniques and philosophies of QDA and TM.  It gives a concise and non-technical overview of various analytic techniques rarely published in one single paper.  It will help encourage collaborations of researchers from different domains by removing the language barrier.  The results support dynamically involving humans with domain-specific knowledge into automated text analysis process, as human input helps extract more specific information than TM alone.  It has reference values for designing semi-automated algorithms to analyse big textual data collected from small-N and large-N samples.  This paper also adds knowledge to a salient but sometimes neglected topic -- the relations between ITs, theory and methods (or methodologies).  Lacking consensus about this topic cannot be specific or uniquely important to QR given that computational approaches are penetrating into all sciences.  This topic deserves immediate research attention to avoid replacing human wisdom with artificial intelligence too soon and without justification.

\sec{Acknowledgements}
The research was developed when the author, ST, was on academic exchange with the University of Nottingham and substantially revised while working at Swansea University.  ST would like to thank the team members of a qualitative study for sharing the primary data, where Dr. Puja Myles was the principal investigator and Dr. Anand Ahankari supervised the data collection and was involved in transcribing/translating the interviews.  Thanks also extend to Dr. Myles and Dr. Jack Gibson of the University of Nottingham and Dr. Petra Buttner of James Cook University for their specialist opinions, late Professor Damon Berridge and Mr. Martin Heaven of Swansea University for being internal reviewers and Dr. Bridie Evans for matching internal reviewers, and Dr. Enrique Garc\'{i}a of the University of Limerick for re-writing suggestions.  ST would also like to thank the hospitality of the Health Protection and Influenza Research Group and Division of Epidemiology and Public Health of the University of Nottingham and their permission to use the data.  Usual disclaimer applies.

\sec{Ethics approval}
The raw data were originally gathered by a project that was approved by the Medical School Research Ethics Committee of the University of Nottingham (reference number: OVS08102013 SoM EPH) and the Institutional Ethics Committee of Maharashtra Association of Anthropological Sciences of India (reference number: MAAS-IEC/2013/001).  No patient data was employed.  The 20 practitioners voluntarily participated in the study.

\sec{Funding statements}
This project was developed when ST was conducting research under the International Postdoctoral Fellowship scheme of the University of Nottingham.

\sec{Completing interests}
None.


\bibliography{ref-addition}

\begin{thebibliography}{48}
\providecommand{\natexlab}[1]{#1}
\providecommand{\url}[1]{\textrm{#1}}
\providecommand{\urlprefix}{ }
\expandafter\ifx\csname urlstyle\endcsname\relax
  \providecommand{\doi}[1]{doi:\discretionary{}{}{}#1}\else
  \providecommand{\doi}{doi:\discretionary{}{}{}\begingroup
  \urlstyle{rm}\Url}\fi
\providecommand{\eprint}[2][]{\url{#2}}
\providecommand{\bibinfo}[2]{#2}
\ifx\xfnm\undefined \def\xfnm[#1]{\unskip,\space#1}\fi
\makeatletter\def\@biblabel#1{#1.}\makeatother
\bibitem[{Winnenburg \emph{et~al.}(2008)Winnenburg, W\"{a}chter, Plake, Doms
  and Schroeder}]{Winnenburgetal2008}
\bibinfo{author}{Winnenburg, R.}, \bibinfo{author}{W\"{a}chter, T.},
  \bibinfo{author}{Plake, C.}, \bibinfo{author}{Doms, A.} and
  \bibinfo{author}{Schroeder, M.} (\bibinfo{year}{2008}) \bibinfo{title}{Facts
  from text: can text mining help to scale-up high-quality manual curation of
  gene products with ontologies?},
  \hspace{0pt}\emph{\bibinfo{journal}{Briefings in Bioinformatics}},
  \bibinfo{volume}{9}(\bibinfo{number}{6}),~\bibinfo{pages}{466}.
\bibitem[{Tonelli(2018)}]{Tonelli2018}
\bibinfo{author}{Tonelli, M.~R.} (\bibinfo{year}{2018})
  \bibinfo{title}{Clinical judgement in precision medicine},
  \hspace{0pt}\emph{\bibinfo{journal}{Journal of evaluation in clinical
  practice}},
  \bibinfo{volume}{24}(\bibinfo{number}{3}),~\bibinfo{pages}{646--648}.
\bibitem[{Kelle(1997)}]{Kelle1997}
\bibinfo{author}{Kelle, U.} (\bibinfo{year}{1997}) \bibinfo{title}{Theory
  Building in Qualitative Research and Computer Programs for the Management of
  Textual Data}, \hspace{0pt}\emph{\bibinfo{journal}{Sociological Research
  Online}}, \bibinfo{volume}{2}(\bibinfo{number}{2}).
\newblock \urlprefix\url{http://www.socresonline.org.uk/2/2/1.html} (last
  accessed \bibinfo{note}{3 February 2020}).
\bibitem[{Palinkas \emph{et~al.}(2015)Palinkas, Horwitz, Green
  \emph{et~al.}}]{Palinkas2015}
\bibinfo{author}{Palinkas, L.~A.}, \bibinfo{author}{Horwitz, S.~M.},
  \bibinfo{author}{Green, C.~A.}, \bibinfo{author}{Wisdom, J.~P.},
  \bibinfo{author}{Duan, N.} and \bibinfo{author}{Hoagwood, K.}
  (\bibinfo{year}{2015}) \bibinfo{title}{Purposeful sampling for qualitative
  data collection and analysis in mixed method implementation research},
  \hspace{0pt}\emph{\bibinfo{journal}{Administration and Policy in Mental
  Health and Mental Health Services Research}},
  \bibinfo{volume}{42}(\bibinfo{number}{5}),~\bibinfo{pages}{533--544}.
\bibitem[{Wiedemann(2013)}]{Wiedemann2013}
\bibinfo{author}{Wiedemann, G.} (\bibinfo{year}{2013}) \bibinfo{title}{Opening
  up to Big Data: Computer-Assisted Analysis of Textual Data in Social
  Sciences}, \hspace{0pt}\emph{\bibinfo{journal}{Forum Qualitative
  Sozialforschung/Forum: Qualitative Social Research}},
  \bibinfo{volume}{14}(\bibinfo{number}{2}).
\bibitem[{Atkisson \emph{et~al.}(2016)Atkisson, Monaghan and
  Brent}]{Atkissonetal2016}
\bibinfo{author}{Atkisson, C.}, \bibinfo{author}{Monaghan, C.} and
  \bibinfo{author}{Brent, E.} (\bibinfo{year}{2016}) \bibinfo{title}{Using
  computational techniques to fill the gap between qualitative data analysis
  and text analytics}, \hspace{0pt}\emph{\bibinfo{journal}{Tijdschrift
  Kwalon}}, \bibinfo{volume}{15}(\bibinfo{number}{3}).
\bibitem[{Densen(2011)}]{Densen2011}
\bibinfo{author}{Densen, P.} (\bibinfo{year}{2011}) \bibinfo{title}{Challenges
  and opportunities facing medical education},
  \hspace{0pt}\emph{\bibinfo{journal}{Transactions of the American Clinical and
  Climatological Association}}, \bibinfo{volume}{122},~\bibinfo{pages}{48}.
\bibitem[{Evers \emph{et~al.}(2011)Evers, Silver, Mruck and
  Peeters}]{Eversetal2011}
\bibinfo{author}{Evers, J.~C.}, \bibinfo{author}{Silver, C.},
  \bibinfo{author}{Mruck, K.} and \bibinfo{author}{Peeters, B.}
  (\bibinfo{year}{2011}) \bibinfo{title}{Introduction to the {KWALON}
  Experiment: Discussions on Qualitative Data Analysis Software by Developers
  and Users}, \hspace{0pt}\emph{\bibinfo{journal}{Forum Qualitative
  Sozialforschung / Forum: Qualitative Social Research}},
  \bibinfo{volume}{12}(\bibinfo{number}{1}).
\bibitem[{Greenhalgh and Taylor(1997)}]{GreenhalghTaylor1997}
\bibinfo{author}{Greenhalgh, T.} and \bibinfo{author}{Taylor, R.}
  (\bibinfo{year}{1997}) \bibinfo{title}{How to read a paper: Papers that go
  beyond numbers (qualitative research)},
  \hspace{0pt}\emph{\bibinfo{journal}{{BMJ}}},
  \bibinfo{volume}{315}(\bibinfo{number}{7110}),~\bibinfo{pages}{740--743}.
\bibitem[{Green and Britten(1998)}]{GreenBritten1998}
\bibinfo{author}{Green, J.} and \bibinfo{author}{Britten, N.}
  (\bibinfo{year}{1998}) \bibinfo{title}{Qualitative research and evidence
  based medicine}, \hspace{0pt}\emph{\bibinfo{journal}{{BMJ}}},
  \bibinfo{volume}{316}(\bibinfo{number}{7139}),~\bibinfo{pages}{1230}.
\bibitem[{Noble and Smith(2015)}]{NobleSmith2015}
\bibinfo{author}{Noble, H.} and \bibinfo{author}{Smith, J.}
  (\bibinfo{year}{2015}) \bibinfo{title}{Issues of validity and reliability in
  qualitative research}, \hspace{0pt}\emph{\bibinfo{journal}{Evidence Based
  Nursing}},
  \bibinfo{volume}{18}(\bibinfo{number}{2}),~\bibinfo{pages}{34--35}.
\bibitem[{Thomas(2006)}]{Thomas2006}
\bibinfo{author}{Thomas, D.~R.} (\bibinfo{year}{2006}) \bibinfo{title}{A
  general inductive approach for analyzing qualitative evaluation data},
  \hspace{0pt}\emph{\bibinfo{journal}{American journal of evaluation}},
  \bibinfo{volume}{27}(\bibinfo{number}{2}),~\bibinfo{pages}{237--246}.
\bibitem[{Lejeune(2011)}]{Lejeune2011}
\bibinfo{author}{Lejeune, C.} (\bibinfo{year}{2011}) \bibinfo{title}{From
  normal business to financial crisis\dots and back again. An illustration of
  the benefits of Cassandre for qualitative analysis},
  \hspace{0pt}\emph{\bibinfo{journal}{Forum: Qualitative Sozialforschung}},
  \bibinfo{volume}{12}(\bibinfo{number}{1}),~\bibinfo{pages}{19}.
\bibitem[{Yu \emph{et~al.}(2011)Yu, Jannasch-Pennell and DiGangi}]{Yuetal2011}
\bibinfo{author}{Yu, C.~H.}, \bibinfo{author}{Jannasch-Pennell, A.} and
  \bibinfo{author}{DiGangi, S.} (\bibinfo{year}{2011})
  \bibinfo{title}{Compatibility between text mining and qualitative research in
  the perspectives of grounded theory, content analysis, and reliability},
  \hspace{0pt}\emph{\bibinfo{journal}{The Qualitative Report}},
  \bibinfo{volume}{16}(\bibinfo{number}{3}),~\bibinfo{pages}{730--744}.
\bibitem[{Gandomi and Haider(2015)}]{GandomiHaider2015}
\bibinfo{author}{Gandomi, A.} and \bibinfo{author}{Haider, M.}
  (\bibinfo{year}{2015}) \bibinfo{title}{Beyond the hype: Big data concepts,
  methods, and analytics}, \hspace{0pt}\emph{\bibinfo{journal}{International
  Journal of Information Management}},
  \bibinfo{volume}{35}(\bibinfo{number}{2}),~\bibinfo{pages}{137--144}.
\bibitem[{Grimmer and Stewart(2013)}]{GrimmerStewart2013}
\bibinfo{author}{Grimmer, J.} and \bibinfo{author}{Stewart, B.~M.}
  (\bibinfo{year}{2013}) \bibinfo{title}{Text as data: The promise and pitfalls
  of automatic content analysis methods for political texts},
  \hspace{0pt}\emph{\bibinfo{journal}{Political analysis}},
  \bibinfo{volume}{21}(\bibinfo{number}{3}),~\bibinfo{pages}{267--297}.
\bibitem[{Gladwin(1989)}]{Gladwin1989}
\bibinfo{author}{Gladwin, C.~H.} (\bibinfo{year}{1989})
  \emph{\bibinfo{title}{Ethnographic decision tree modeling}}
  Vol.~\bibinfo{volume}{19}.
\newblock \bibinfo{publisher}{Sage}.
\bibitem[{Mason(2010)}]{Mason2010}
\bibinfo{author}{Mason, M.} (\bibinfo{year}{2010}) \bibinfo{title}{Sample size
  and saturation in PhD studies using qualitative interviews},
  \hspace{0pt}\emph{\bibinfo{journal}{Forum qualitative Sozialforschung/Forum:
  qualitative social research}}, \bibinfo{volume}{11}(\bibinfo{number}{3}).
\bibitem[{Ahankari \emph{et~al.}(2017)Ahankari, Myles, Tsang
  \emph{et~al.}}]{Ahankarietal2017}
\bibinfo{author}{Ahankari, A.~S.}, \bibinfo{author}{Myles, P.~R.},
  \bibinfo{author}{Tsang, S.}, \bibinfo{author}{Khan, F.},
  \bibinfo{author}{Atre, S.}, \bibinfo{author}{Langley, T.},
  \bibinfo{author}{Kudale, A.} and \bibinfo{author}{Bains, M.}
  (\bibinfo{year}{2017}) \bibinfo{title}{A qualitative study exploring factors
  influencing clinical decision-making for influenza-like illness in Solapur
  city, Maharashtra, India}, \hspace{0pt}\emph{\bibinfo{journal}{Anthropology
  \& medicine}},~\bibinfo{pages}{1--22}.
\bibitem[{{R Core Team}(2015)}]{Rcore2015}
\bibinfo{author}{{R Core Team}} (\bibinfo{year}{2015}) \emph{\bibinfo{title}{R:
  A Language and Environment for Statistical Computing}}.
\newblock \bibinfo{address}{Vienna, Austria}: \bibinfo{organization}{R
  Foundation for Statistical Computing}.
\newblock \urlprefix\url{https://www.R-project.org/}.
\bibitem[{Feinerer \emph{et~al.}(2008)Feinerer, Hornik and
  Meyer}]{Feinereretal2008}
\bibinfo{author}{Feinerer, I.}, \bibinfo{author}{Hornik, K.} and
  \bibinfo{author}{Meyer, D.} (\bibinfo{year}{2008}) \bibinfo{title}{Text
  Mining Infrastructure in R}, \hspace{0pt}\emph{\bibinfo{journal}{Journal of
  Statistical Software}},
  \bibinfo{volume}{25}(\bibinfo{number}{1}),~\bibinfo{pages}{1--54}.
\bibitem[{Huang(2016)}]{Huang2016}
\bibinfo{author}{Huang, R.} (\bibinfo{year}{2016}) \emph{\bibinfo{title}{RQDA:
  R-based Qualitative Data Analysis}}.
\newblock \urlprefix\url{http://rqda.r-forge.r-project.org/}.
\bibitem[{Csardi and Nepusz(2006)}]{CsardiNepusz2006}
\bibinfo{author}{Csardi, G.} and \bibinfo{author}{Nepusz, T.}
  (\bibinfo{year}{2006}) \bibinfo{title}{The igraph software package for
  complex network research},
  \hspace{0pt}\emph{\bibinfo{journal}{InterJournal}}, \bibinfo{volume}{Complex
  Systems},~\bibinfo{pages}{1695}.
\bibitem[{Mueller(2019)}]{Mueller2019}
\bibinfo{author}{Mueller, A.} (\bibinfo{year}{2019}).
\newblock \emph{\bibinfo{title}{WordCloud}}.
\newblock \bibinfo{organization}{Columbia University}.
\newblock \bibinfo{address}{New York}, \bibinfo{edition}{Python package version
  1.6.0} edn.
\newblock \urlprefix\url{https://amueller.github.io/word_cloud/index.html}.
\newblock (last accessed \bibinfo{note}{12 January 2020}).
\bibitem[{Van~Rossum and Drake(2009)}]{VanRossumDrake2009}
\bibinfo{author}{Van~Rossum, G.} and \bibinfo{author}{Drake, F.~L.}
  (\bibinfo{year}{2009}) \emph{\bibinfo{title}{Python 3 Reference Manual}}.
\newblock \bibinfo{address}{Scotts Valley, CA}:
  \bibinfo{publisher}{CreateSpace}.
\newblock ISBN \bibinfo{isbn}{1441412697}.
\bibitem[{Rinker(2013)}]{Rinker2013}
\bibinfo{author}{Rinker, T.~W.} (\bibinfo{year}{2013})
  \emph{\bibinfo{title}{{qdapDictionaries}: Dictionaries to Accompany the qdap
  Package}}.
\newblock \bibinfo{address}{Buffalo, New York}:
  \bibinfo{organization}{University at Buffalo/SUNY}.
\newblock \urlprefix\url{http://github.com/trinker/qdapDictionaries}.
\bibitem[{Hopkins and King(2010)}]{HopkinsKing2010}
\bibinfo{author}{Hopkins, D.~J.} and \bibinfo{author}{King, G.}
  (\bibinfo{year}{2010}) \bibinfo{title}{A method of automated nonparametric
  content analysis for social science},
  \hspace{0pt}\emph{\bibinfo{journal}{American Journal of Political Science}},
  \bibinfo{volume}{54}(\bibinfo{number}{1}),~\bibinfo{pages}{229--247}.
\bibitem[{Barbour(2001)}]{Barbour2001}
\bibinfo{author}{Barbour, R.~S.} (\bibinfo{year}{2001})
  \bibinfo{title}{Checklists for improving rigour in qualitative research: a
  case of the tail wagging the dog?},
  \hspace{0pt}\emph{\bibinfo{journal}{{BMJ}}},
  \bibinfo{volume}{322}(\bibinfo{number}{7294}),~\bibinfo{pages}{1115--1117}.
\bibitem[{Rowley(2012)}]{Rowley2012}
\bibinfo{author}{Rowley, J.} (\bibinfo{year}{2012}) \bibinfo{title}{Conducting
  research interviews}, \hspace{0pt}\emph{\bibinfo{journal}{Management research
  review}},
  \bibinfo{volume}{35}(\bibinfo{number}{3/4}),~\bibinfo{pages}{260--271}.
\bibitem[{Newman and Girvan(2004)}]{NewmanGirvan2004}
\bibinfo{author}{Newman, M.~E.} and \bibinfo{author}{Girvan, M.}
  (\bibinfo{year}{2004}) \bibinfo{title}{Finding and evaluating community
  structure in networks}, \hspace{0pt}\emph{\bibinfo{journal}{Physical review
  E}}, \bibinfo{volume}{69}(\bibinfo{number}{2}),~\bibinfo{pages}{026113}.
\bibitem[{Cicchetti(1994)}]{Cicchetti1994}
\bibinfo{author}{Cicchetti, D.~V.} (\bibinfo{year}{1994})
  \bibinfo{title}{Guidelines, criteria, and rules of thumb for evaluating
  normed and standardized assessment instruments in psychology.},
  \hspace{0pt}\emph{\bibinfo{journal}{Psychological assessment}},
  \bibinfo{volume}{6}(\bibinfo{number}{4}),~\bibinfo{pages}{284}.
\bibitem[{Steiger(1980)}]{Steiger1980}
\bibinfo{author}{Steiger, J.~H.} (\bibinfo{year}{1980}) \bibinfo{title}{Tests
  for comparing elements of a correlation matrix},
  \hspace{0pt}\emph{\bibinfo{journal}{Psychological bulletin}},
  \bibinfo{volume}{87}(\bibinfo{number}{2}),~\bibinfo{pages}{245--251}.
\bibitem[{Feldman \emph{et~al.}(2016)Feldman, Hazekamp and
  Chawla}]{Feldmanetal2016}
\bibinfo{author}{Feldman, K.}, \bibinfo{author}{Hazekamp, N.} and
  \bibinfo{author}{Chawla, N.~V.} (\bibinfo{year}{2016}) \bibinfo{title}{Mining
  the clinical narrative: all text are not equal}.
\newblock In \emph{\bibinfo{booktitle}{Healthcare Informatics (ICHI), 2016 IEEE
  International Conference on}}, \hspace{0pt}pp. \bibinfo{pages}{271--280}.
\newblock \bibinfo{organization}{IEEE}.
\bibitem[{Sutton and Austin(2015)}]{SuttonAustin2015}
\bibinfo{author}{Sutton, J.} and \bibinfo{author}{Austin, Z.}
  (\bibinfo{year}{2015}) \bibinfo{title}{Qualitative research: data collection,
  analysis, and management}, \hspace{0pt}\emph{\bibinfo{journal}{The Canadian
  journal of hospital pharmacy}},
  \bibinfo{volume}{68}(\bibinfo{number}{3}),~\bibinfo{pages}{226}.
\bibitem[{Hutton \emph{et~al.}(2015)Hutton, Salanti, Caldwell
  \emph{et~al.}}]{Huttonetal2015}
\bibinfo{author}{Hutton, B.}, \bibinfo{author}{Salanti, G.},
  \bibinfo{author}{Caldwell, D.~M.}, \bibinfo{author}{Chaimani, A.},
  \bibinfo{author}{Schmid, C.~H.}, \bibinfo{author}{Cameron, C.},
  \bibinfo{author}{Ioannidis, J.~P.}, \bibinfo{author}{Straus, S.},
  \bibinfo{author}{Thorlund, K.}, \bibinfo{author}{Jansen, J.~P.} \emph{et~al.}
  (\bibinfo{year}{2015}) \bibinfo{title}{The PRISMA extension statement for
  reporting of systematic reviews incorporating network meta-analyses of health
  care interventions: checklist and explanations},
  \hspace{0pt}\emph{\bibinfo{journal}{Annals of internal medicine}},
  \bibinfo{volume}{162}(\bibinfo{number}{11}),~\bibinfo{pages}{777--784}.
\bibitem[{Agresti \emph{et~al.}(2016)Agresti, Franklin and
  Klingenberg}]{AgrestiChristine2016}
\bibinfo{author}{Agresti, A.}, \bibinfo{author}{Franklin, C.~A.} and
  \bibinfo{author}{Klingenberg, B.} (\bibinfo{year}{2016})
  \emph{\bibinfo{title}{Statistics: The art and science of learning from
  data}}.
\newblock \bibinfo{publisher}{Pearson}.
\bibitem[{Efron and Tibshirani(1994)}]{EfronTibshirani1994}
\bibinfo{author}{Efron, B.} and \bibinfo{author}{Tibshirani, R.~J.}
  (\bibinfo{year}{1994}) \emph{\bibinfo{title}{An introduction to the
  bootstrap}}.
\newblock \bibinfo{publisher}{CRC press}.
\bibitem[{Grimmer(2015)}]{Grimmer2015}
\bibinfo{author}{Grimmer, J.} (\bibinfo{year}{2015}) \bibinfo{title}{We are all
  social scientists now: how big data, machine learning, and causal inference
  work together}, \hspace{0pt}\emph{\bibinfo{journal}{PS: Political Science \&
  Politics}},
  \bibinfo{volume}{48}(\bibinfo{number}{1}),~\bibinfo{pages}{80--83}.
\bibitem[{Ryan and Bernard(2006)}]{RyanBernard2006}
\bibinfo{author}{Ryan, G.~W.} and \bibinfo{author}{Bernard, H.~R.}
  (\bibinfo{year}{2006}) \bibinfo{title}{Testing an ethnographic decision tree
  model on a national sample: Recycling beverage cans},
  \hspace{0pt}\emph{\bibinfo{journal}{Human Organization}},
  \bibinfo{volume}{65}(\bibinfo{number}{1}),~\bibinfo{pages}{103--114}.
\bibitem[{Rayson \emph{et~al.}(2004)Rayson, Berridge and
  Francis}]{Raysonetal2004}
\bibinfo{author}{Rayson, P.}, \bibinfo{author}{Berridge, D.} and
  \bibinfo{author}{Francis, B.} (\bibinfo{year}{2004})
  \bibinfo{title}{Extending the Cochran rule for the comparison of word
  frequencies between corpora}.
\newblock In \emph{\bibinfo{booktitle}{7th International Conference on
  Statistical analysis of textual data (JADT 2004)}}, \hspace{0pt}pp.
  \bibinfo{pages}{926--936}.
\bibitem[{Strauss \emph{et~al.}(1990)Strauss, Corbin
  \emph{et~al.}}]{StraussCorbin1990}
\bibinfo{author}{Strauss, A.~L.}, \bibinfo{author}{Corbin, J.~M.} \emph{et~al.}
  (\bibinfo{year}{1990}) \emph{\bibinfo{title}{Basics of qualitative research}}
  Vol.~\bibinfo{volume}{15}.
\newblock \bibinfo{publisher}{Sage Newbury Park, CA}.
\bibitem[{Thorne(2000)}]{Thorne2000}
\bibinfo{author}{Thorne, S.} (\bibinfo{year}{2000}) \bibinfo{title}{Data
  analysis in qualitative research},
  \hspace{0pt}\emph{\bibinfo{journal}{Evidence based nursing}},
  \bibinfo{volume}{3}(\bibinfo{number}{3}),~\bibinfo{pages}{68--70}.
\bibitem[{Lunny \emph{et~al.}(2016)Lunny, McKenzie and
  McDonald}]{Lunnyetal2016}
\bibinfo{author}{Lunny, C.}, \bibinfo{author}{McKenzie, J.~E.} and
  \bibinfo{author}{McDonald, S.} (\bibinfo{year}{2016})
  \bibinfo{title}{Retrieval of overviews of systematic reviews in MEDLINE was
  improved by the development of an objectively derived and validated search
  strategy}, \hspace{0pt}\emph{\bibinfo{journal}{Journal of clinical
  epidemiology}}, \bibinfo{volume}{74},~\bibinfo{pages}{107--118}.
\bibitem[{McAlister \emph{et~al.}(2017)McAlister, Lee, Ehlert
  \emph{et~al.}}]{Mcalisteretal2017}
\bibinfo{author}{McAlister, A.}, \bibinfo{author}{Lee, D.},
  \bibinfo{author}{Ehlert, K.}, \bibinfo{author}{Kajfez, R.},
  \bibinfo{author}{Faber, C.} and \bibinfo{author}{Kennedy, M.}
  (\bibinfo{year}{2017}) \bibinfo{title}{Qualitative coding: An approach to
  assess inter-rater reliability}.
\newblock In \emph{\bibinfo{booktitle}{ASEE Annual Conference \& Exposition,
  Columbus, Ohio}}, \hspace{0pt}\urlprefix\url{https://peer. asee. org/28777}.
\bibitem[{Armstrong \emph{et~al.}(1997)Armstrong, Gosling, Weinman and
  Marteau}]{Armstrongetal1997}
\bibinfo{author}{Armstrong, D.}, \bibinfo{author}{Gosling, A.},
  \bibinfo{author}{Weinman, J.} and \bibinfo{author}{Marteau, T.}
  (\bibinfo{year}{1997}) \bibinfo{title}{The place of inter-rater reliability
  in qualitative research: an empirical study},
  \hspace{0pt}\emph{\bibinfo{journal}{Sociology}},
  \bibinfo{volume}{31}(\bibinfo{number}{3}),~\bibinfo{pages}{597--606}.
\bibitem[{Moret \emph{et~al.}(2007)Moret, Reuzel, Van Der~Wilt and
  Grin}]{Moretetal2007}
\bibinfo{author}{Moret, M.}, \bibinfo{author}{Reuzel, R.}, \bibinfo{author}{Van
  Der~Wilt, G.~J.} and \bibinfo{author}{Grin, J.} (\bibinfo{year}{2007})
  \bibinfo{title}{Validity and reliability of qualitative data analysis:
  Interobserver agreement in reconstructing interpretative frames},
  \hspace{0pt}\emph{\bibinfo{journal}{Field Methods}},
  \bibinfo{volume}{19}(\bibinfo{number}{1}),~\bibinfo{pages}{24--39}.
\bibitem[{Shuval \emph{et~al.}(2011)Shuval, Harker, Roudsari
  \emph{et~al.}}]{Shuvaletal2011}
\bibinfo{author}{Shuval, K.}, \bibinfo{author}{Harker, K.},
  \bibinfo{author}{Roudsari, B.}, \bibinfo{author}{Groce, N.~E.},
  \bibinfo{author}{Mills, B.}, \bibinfo{author}{Siddiqi, Z.} and
  \bibinfo{author}{Shachak, A.} (\bibinfo{year}{2011}) \bibinfo{title}{Is
  Qualitative Research Second Class Science? A Quantitative Longitudinal
  Examination of Qualitative Research in Medical Journals},
  \hspace{0pt}\emph{\bibinfo{journal}{Plos One}},
  \bibinfo{volume}{6}(\bibinfo{number}{2}),~\bibinfo{pages}{e16937}.
\bibitem[{Pope \emph{et~al.}(2000)Pope, Ziebland and Mays}]{Popeetal2000}
\bibinfo{author}{Pope, C.}, \bibinfo{author}{Ziebland, S.} and
  \bibinfo{author}{Mays, N.} (\bibinfo{year}{2000}) \bibinfo{title}{Analysing
  qualitative data}, \hspace{0pt}\emph{\bibinfo{journal}{{BMJ}}},
  \bibinfo{volume}{320}(\bibinfo{number}{7227}),~\bibinfo{pages}{114--116}.

\end{thebibliography}
\bibliographystyle{elsarticle-nbr-jecp}


\setcounter{figure}{0}
\renewcommand{\thefigure}{A\arabic{figure}}
\newgeometry{top=2.54cm,bottom=2.54cm,left=2.54cm,right=2.54cm}
\begin{landscape}
\sec{Appendix}
  \begin{figure}[H]
 \centering
  \includegraphics[trim=200 250 250 200,clip,scale=0.33]{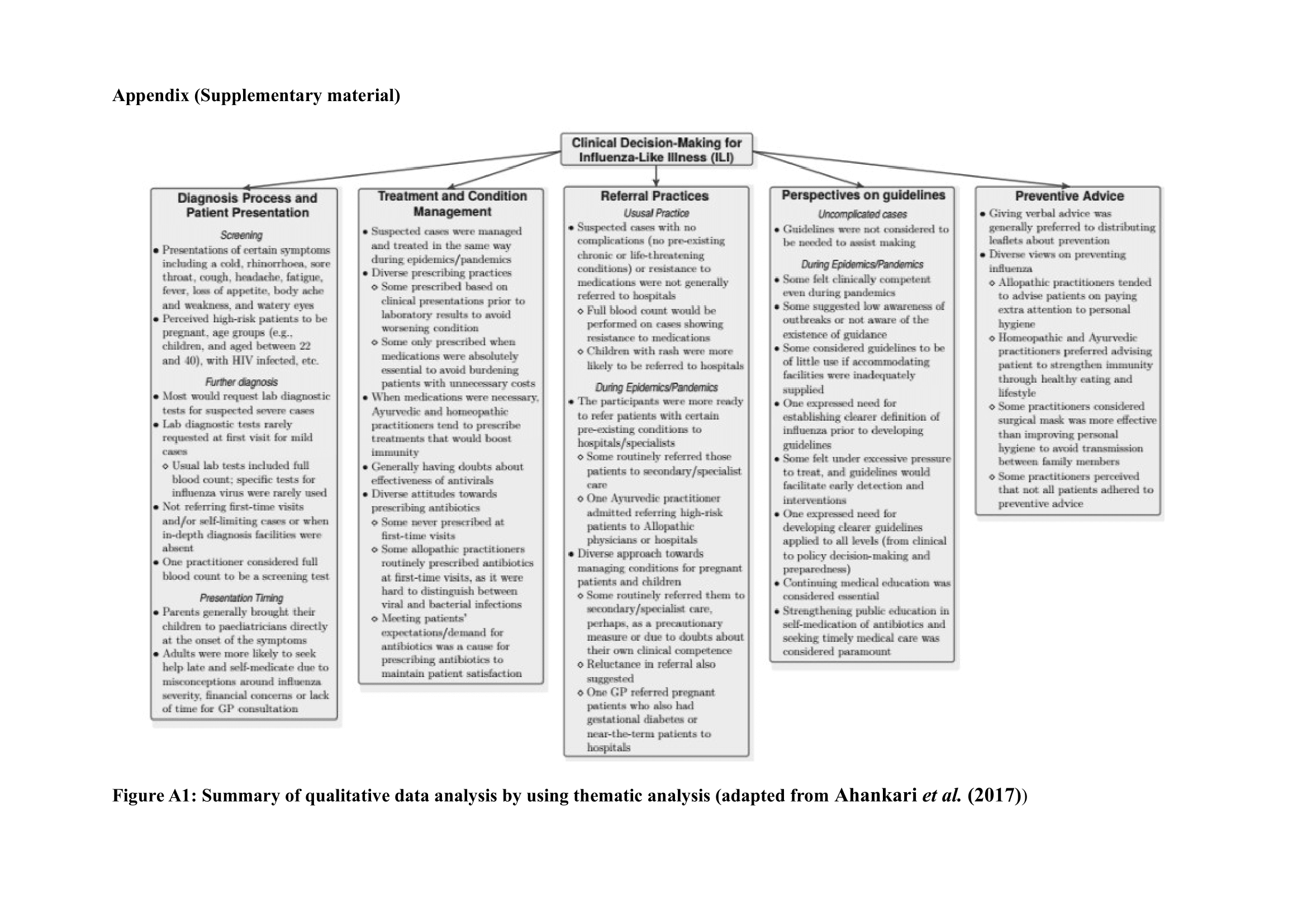}
  \caption{Summary of qualitative data analysis by using thematic analysis (adapted from \citet*{Ahankarietal2017})}
  \label{figure-a1}
  \end{figure}
\end{landscape}

\end{document}